\documentclass{PoS}
\usepackage{multirow}
\usepackage{enumerate}
\usepackage{amsmath,amssymb,marvosym,soul}
\usepackage{slashed,fancybox,empheq,accents}
\usepackage{graphicx}
\allowdisplaybreaks

\title{ICHEP2012 Physics Highlights}

\ShortTitle{ICHEP2012 Highlights}

\author{{Riccardo Barbieri}\\
        Scuola Normale Superiore and INFN, Pisa, Italy\\
        E-mail: \email{barbieri@sns.it}}

\abstract{I describe some of the highlights of ICHEP2012 with an eye to the current status of particle physics.}

\FullConference{36th International Conference on High Energy Physics,\\
		July 4-11, 2012\\
		Melbourne, Australia}

\begin{document}

\section{The synthetic nature of Particle Physics}

To try to summarize a conference like ICHEP2012 an organization principle is needed. To this end I think there is hardly a better way than to refer to the Lagrangian of the Standard Model, suitably extended to describe neutrino masses\footnote{The notation is concise but precise as well. In particular $\Psi$ is a multiplet that contains all matter fermions, including the three right-handed neutrinos $N_i$, one per generation, $i=1, 2,3$.}
\begin{align}
\mathcal{L}_{SM} = & -\frac{1}{4} F^a_{\mu\nu}F^{a \mu\nu} + i \bar{\Psi}\slashed{D}\Psi \,
& {\text the~gauge~sector} \,
\\
 & +|D_\mu \phi|^2 - V(\phi) \,
&  {\text the~ symmetry~breaking~ sector} \,
\label{EWSB}
\\
 & + \Psi_i \lambda_{ij} \Psi_j \phi \,
&  {\text the~ flavour~ sector} \,
\label{flavour}
\\
 & +N_i M_{ij} N_j \,
&  {\text the~ neutrino~mass~ sector~ (if~Majorana)} \,
\end{align}
That this Lagrangian be apt to describe all known interactions among the elementary particles (except the ultra-feeble gravity) illustrates the main character of Particle Physics: its synthetic nature, which I think is the true meaning of reductionism.
This can be said although without neglecting that sooner or later one would like to incorporate in a synthetic description of nature at least two other facts: i) the existence of Dark Matter; ii) the origin of the  matter-antimatter asymmetry in the universe, particularly tiny, although non zero, if one goes back in cosmological  time.

Needless to say, with the announcement of the discovery of a Higgs-like particle \cite{:2012gk,:2012gu}, the main focus of ICHEP2012 has been on the electroweak symmetry breaking sector. It would be inappropriate, however, not to mention also some of the results that have emerged in the gauge sector, in the neutrino sector and in the flavour sector, as  briefly done below in this order.

\section{The gauge sector \cite{Dixon,Campbell}}

A major pending question in the gauge sector is whether it will ever be possible to solve QCD. One is certainly not there yet. In this broad context, nevertheless, I find useful to underline a cross talk that is taking place since about two decades between formal considerations inspired by string theory or N=4 super-Yang-Mills theories and {\it hard} QCD calculations needed in LHC analyses. This has led to a variety of results in different directions, but a most remarkable one is the progressive development  and refinement of the so called {\it maximal unitarity} method, which allows  to make NLO computations of  processes of increasing complexity, like
\begin{align}
pp &\rightarrow  Z, W, \gamma^* {\text + N jets}~,&
pp&\rightarrow  t \bar{t} {\text + N jets}~,
\end{align}
with $N$ up to 5. These methods have become of current use and are normally interfaced with MonteCarlo programs for a realistic description of the final states. First steps towards an extension of these methods to NNLO calculations are being made \cite{Kosower}.

\section{Neutrinos \cite{Kobayashi,Cao,Gonzalez}}

At least from an experimental point of view a pretty clear list of open questions in neutrino physics, ordered in importance to my own taste, is:
\begin{itemize}
\item 1. Are neutrinos of Majorana or Dirac type?
\item 2. Which is the value of $\theta_{13}$?
\item 3. Which is the c.o.m. of the neutrinos? In particular is their spectrum {\it normal} of {\it inverted}?
\item 4. Is there CP violation in the leptons as well?
\item 5. How many neutrinos there are: 3 or more?
\end{itemize}
The important new result is that the answer to the second question above is now known with significant precision,  due to data from short base-line reactor experiments, mostly Daya Bay but also RENO and Double Chooz, with some additional information from increased statistics of long-baseline experiments,
T2K  and MINOS. The precise determination of $\theta_{13}$ can only be made by statistically combining the results of all the neutrino oscillation searches, within a three neutrino hypothesis, and depends on the choice of reactor 
fluxes: a fair value is 
\begin{equation}
\sin^2{\theta_{13}} = 0.024 \pm 0.003,
\end{equation}
i.e. $\theta_{13}\approx 8^0\div 9^0$. The same overall oscillation fit, also slightly dependent on the chosen ordering of the neutrino spectrum, normal or inverted, indicates a "non-maximal" value for  $\theta_{23}$, either below or above $45^0$ by about 5 degrees. Except for the CP violating phase, all the oscillation parameters, i.e. the two independent squared mass differences and the three $\sin^2{\theta_{ij}}$, are now known with a relative precision that ranges from about 2 to 15$\%$.
What to make out of these numbers is a matter of intense discussion.

\section{Flavour \cite{Kuno,Perez,Stone,Tarantino}}

As it will be elaborated in a while, a central question in flavour physics is how to interpret the mounting empirical evidence for the Cabibbo Kobayashi Maskawa (CKM) picture of flavour and CP violation. Before doing that, however, it is necessary to enumerate and  briefly comment on some potential {\it exceptions} that have shown up recently (or even not so recently).

Much discussion has arisen about the emerging evidence for direct CP violation in charm decays from combined CDF\cite{Tonelli} and LHCb \cite{Garra}  results (HFAG average)
\begin{equation}
\Delta A_{CP}(D\rightarrow P P) \equiv \Delta A_{CP}(K^+K^-) -\Delta A_{CP}(\pi^+ \pi^-) = (- 6.45 \pm 1.80) 10^{-3}
\label{CPVinD}
\end{equation}
Is it {\it old} or {\it new} physics? Remember the only partially analogous discussion made long ago about direct CP violation in Kaon decays (the $\epsilon^\prime$ parameter). One would think that charm decays are easier to predict. This does not look to be the case, however, at the level of the observation in (\ref{CPVinD}) \cite{Soni}. Standard theoretical tools used in this kind of analyses (factorization,  approximate $SU(3)$, etc.) are problematic, if not in contrast with the data. More data on different D-decay channels could help to resolve the issue, but, for the time being, the judgement is suspended.

  BaBar has presented measurements of $B\rightarrow D + \tau +\nu$ and $B\rightarrow D^* + \tau +\nu$, which, normalized to the same decays into electrons or muons and fitted together, deviate from the SM values by slightly more than $3\sigma$ \cite{DeNardo}. A charged Higgs boson exchange at tree level would not be capable of accounting for this putative discrepancy in both the $D$ and the $D^*$ cases simultaneously. If confirmed, one would be confronted with a deviation from the SM in a {\it large} tree level decay, not easy to digest. 
    
 A problem in $B$ to $\tau$ purely leptonic decays exists in fact since quite some  time, with a measured Branching Ratio by BaBar and Belle  higher than the SM expectation. A recent combined Belle result \cite{Yook}
 \begin{equation}
 BR(B\rightarrow \tau \nu) = (0.96\pm 0.26) 10^{-4}
 \end{equation}
 is significantly lower than the previous ones, thus apparently ameliorating the agreement with the SM. It must be said, however,  that the 
 SM prediction rests on the chosen value of $V_{ub}$, which suffers of the well known discrepancy between {\it inclusive} and {\it exclusive} determinations. At least this last problem remains open \cite{Nakao,Urquijo}.

A last  problem  that one may want to mention is the measurement of the forward-backward $t \bar{t}$ asymmetry at the Tevatron with the combination of CDF  \cite{Grohsjean} and D0  \cite{Hays} results summarized in Table 1,  together with the SM expectation. Before any firm conclusion can be drawn,  it is evident that an assessment of the SM uncertainty is necessary, although foreseen to be small. It is debatable how far can the LHC go in trying to resolve the issue. It must also be said that models trying to accommodate these results do not have an easy life in coping with a body of different measurements in top physics and beyond.

\begin{table}[tbp]
\renewcommand{\arraystretch}{1.3}
\centering
\begin{tabular}{ccc}
\hline
% mass & 
&Exp (CDF + D0) & SM(QCD + EW)\\
\hline
$A^{inc}_{FB}$ & $ \approx (18\pm 4)\%$ & $\approx (6.6 \pm ?)\%$\\
$A^{> 450 GeV}_{FB}$ & $ \approx (28\pm 6)\%$ & $\approx (10 \pm ?)\%$\\
\hline
\end{tabular}
\caption{$p \bar{p} \rightarrow t \bar{t} $ forward-backward asymmetry averaged over  any $t\bar{t}$ invariant mass (inc) and above 450 GeV \cite{Perez}}
\label{tab:bounds-ew}
\end{table}

These issues notwithstanding, the clearly mounting experimental evidence in favour of the CKM picture of flavour physics poses a neat problem, as alluded to at the beginning of this Section. How is this compatible with the view that requires new physics at the Fermi scale, $G_F^{-1/2}$, to explain its naturalness, as discussed below? This problem is quantitatively expressed by the very strong lower bounds on the scale, much above $G_F^{-1/2}$,  associated with flavour-breaking  beyond-the-SM operators that affect key observables like the $\epsilon$ parameter in $K$ physics. The novelty here is that the list of such operators, and of the corresponding flavour observables, has been very significantly expanding in the last years.  On the other hand, from a theory point of view, it is a fact that models that try to account for a natural Fermi scale  generally introduce new degrees of freedom carrying flavour indices, whose exchanges affect these flavour observables.

A problem like this, if  the background is correct, calls for a   suitable mechanism that operates to keep these extra effects under  control. Such mechanism can have a dynamical or a symmetry origin, or both. To me it is suggestive to consider the flavour symmetry exhibited by the quark sector of the SM in the limit in which one neglects the masses of the first two generations and their small communication with the third family. Formally it is a $U(2)^3$ symmetry acting on the first two generations of left handed electroweak doublets, $q^i_L$, and on the right handed electroweak singlets, $u^i_R$ and $d^i_R$, $i=1,2$. A symmetry like this, broken along suitable directions defined by
\begin{align}
V&\sim({\bold 2},{\bold 1},{\bold 1}),& \Delta_u&\sim({\bold 2},{\bold 2},{\bold 1}),& \Delta_d&\sim({\bold 2},{\bold 1},{\bold 2}),
\label{MU2spurions}
\end{align}
as minimally required for a realistic description of masses and mixings of all quarks, is barely capable to explain in some concrete models of ElectroWeak Symmetry Breaking (EWSB) why no deviation from the CKM picture of flavour physics has been observed so far. Conversely, some of the current searches of new phenomena in flavour physics appear strongly motivated, both in B- and in K-physics. There is no need to say how much a scenario like this would be welcome if one wants to have some clue to attack the puzzle of flavour.

\section{ElectroWeak Symmetry Breaking}

The discovery of a Higgs-like particle is the event of ICHEP2012 or in fact of Physics {\it tout court} in many years.
In this case the key question is: Should one view this discovery as the coronation of the SM or a major milestone along a path yet largely unexplored?

Let us first put the discovery in prospective, giving for granted that the data to come will definitely establish the relation of the newly found particle to EWSB, e.g. by measuring, among other things, its spin and parity. Overwhelming experimental evidence exists since quite some time for the spontaneous breaking of the electroweak symmetry. This entails the presence of three Goldstone bosons - often called $\pi_a, a=1,2,3$, in analogy with the pions of QCD - 
which provide the longitudinal degrees of freedom of the $W$ and $Z$ bosons, as needed for them to get a mass \cite{Englert:1964et,Higgs:1964pj}.
The newly found particle, $h$, is very likely to complete with these "pions" a quartet of real scalar fields
\begin{equation}
\phi = (\pi_a, h)
\end{equation}
or a doublet of complex fields, which form a linearly transforming, complete multiplet of the gauge $SU(2)\times U(1)$ symmetry. This is the field $\phi$ of equation (\ref{EWSB}), whose forth component LHC has discovered. Rather LHC has discovered the quantum excitation of $h$, whose classical component has a non vanishing value $v$, uniformly spread over all space-time, as a consequence of the self-interactions of $\phi$.

Needless to say all this is contained, and as such foreseen, in the Lagrangian of the SM. The same SM Lagrangian dictates the couplings of $\phi$,  some of which are important in the successful fit of the ElectroWeak Precision Tests (EWPT) for $m_h = 90^{+30}_{-23}$ GeV \cite{Baak:2012kk}.
Yet the question set in the first paragraph of this Section appears more motivated then ever. Now that $\phi$ is known to be a piece of the physical reality,  two puzzles associated with its interactions, the flavour puzzle in $\Psi_i \lambda_{ij} \Psi_j \phi$, eq. (\ref{flavour}), and the hierarchy puzzle in $V(\phi)$, eq. (\ref{EWSB}),  become more pressing then they have ever been.

\subsection{The interactions of $\phi$ \cite{Hawkings,Incandela,Shalhout,Pomarol}}

A first step consists in measuring the couplings of $h$ to all known particles, and to itself. Unlike the case of the   SM, in which these couplings are determined   by the mass of the particle  which $h$ is coupled to, spontaneous breaking of the 
electroweak gauge symmetry can be described, at least formally, while leaving most of them free. A motivated selection and definition of these couplings is
given by the effective Lagrangian (after EWSB)
\begin{equation}
\mathcal{L}^{eff} = \mathcal{L}^{h-vectors (tree)} + \mathcal{L}^{h-fermions} + \mathcal{L}^{h-vectors (loops)}
\label{L-couplings}
\end{equation}
where
\begin{equation}
\mathcal{L}^{h-vectors (tree)}= c_V 
(\frac{2 m_W^2}{v} W_\mu^+ W_\mu^- +
\frac{ m_Z^2}{v} Z_\mu^2) h 
\label{tree}
\end{equation}

\begin{equation}
 \mathcal{L}^{h-fermions}= c_t \frac{m_t}{v} \bar{t} t h + c_b ( \frac{m_b}{v} \bar{b} b + \frac{m_\tau}{v} \bar{\tau} \tau) h 
 \label{fermions}
\end{equation}

\begin{equation}
 \mathcal{L}^{h-vectors (loops)} = - 0.81 c_\gamma \frac{\alpha}{\pi v}
F_{\mu\nu} F^{\mu\nu} h +
0.97 c_g \frac{\alpha_S}{12\pi v}
G_{\mu\nu} G^{\mu\nu} h .
\label{loops}
\end{equation}
In the SM all these dimensionless coefficients are equal to 1 with sufficient accuracy. The two couplings arising from 
loops in (\ref{loops}) depend, in turn, on the tree level couplings $c_V, c_t, c_b$ via
\begin{align}
c_g &= 1.06 c_t - 0.06 c_b + \delta c_g~,&
c_\gamma& = - 0.28 c_t + 1.28 c_V + \delta c_\gamma~,
\end{align}
with the residual $ \delta c_g$ and $ \delta c_\gamma$ that would have to come from loops of new particles and vanish therefore in the SM.

Several sources can be identified for deviations of these couplings from their SM value:
\begin{itemize}
\item The doublet $\phi$ could be accompanied by other scalars (like a singlet or a second doublet) also acquiring a vev, which could induce tree level changes in $c_V, c_t, c_b$;
\item $\phi$ could be a "composite" particle originating from a new nearby  strong interaction, with possible effects on all the couplings above;
\item New particles in loops could significantly affect $c_g$ and $c_\gamma$.
\end{itemize}
To limit the number of parameters, eqs. (\ref{tree}) to (\ref{loops}) incorporate some inputs, defendable at different levels: custodial symmetry, which maintains the SM ratio of the  tree level couplings to the W and the Z, and the relative couplings to $\tau$ and bottom still related to their mass ratio, suggested by the fact that $\bar{b} b$ and $\bar{\tau} \tau$ have the same hypercharge, different from the one of $\bar{t} t$. 
Other couplings are not included in (\ref{L-couplings}), like the loop-induced coupling to the $Z \gamma$, the coupling to the muon or, last but not least, the self-couplings of $h$, since their measurement will have to wait for a major step in the luminosity of the LHC. It is also possible that the Higgs boson has a coupling to an "invisible" sector, perhaps constituting (a part of) the Dark Matter in the universe.

Educated guesses for the uncertainties that will affect the determination of these couplings range from $5$ to $15\%$ with 300 $fb^{-1}$ of integrated luminosity at $\sqrt{s} = 14$ TeV of the LHC\footnote{This assumes a conceivable reduction in the present  theoretical uncertainties as well.}: a major step, relative to the current situation. How significant will it be, compared to the exploring power of the direct searches in similar conditions of the collider? This is an important question that may deserve further study. In a weekly coupled picture of EWSB, decently motivated, I find hard to think of a  deviation, say at 10$\%$ level,  in some of the couplings above without any signal observed in direct searches. The case may be more open  in a strongly coupled theory of EWSB. 

Other than the precise measurements of the Higgs boson couplings, the search for more Higgs bosons is also important. To set a precise strategy for this search is not easy. In the case of a neutral Higgs boson, other than extending to regions of higher mass the channels studied so far, new channels of interest are $H\rightarrow h h, a a$ and $H\rightarrow a Z$, where $h$ is the newly found particle and $a$ is a pseudoscalar decaying into $b\bar{b},~\tau \bar{\tau}$, not easily accessible in single production.

\subsection{Is $\phi$ natural? \cite{Pomarol, Sundrum}}

The issue of  {\it naturalness} of the Higgs boson mass is with us since the seventies, at least when the {\it hierarchy} problem was first discussed in the context of GUTs. This may have led somebody to think that the naturalness problem is a idle theoretical concept. To me this is a wrong  thought. On the contrary the naturalness problem has its roots in the very way field theory has been systematically applied to subatomic physics so far to implement the intuitive notion of separation of  different physical scales. Specifically, a natural Fermi scale would make understandable why the great empirical success of the SM does not depend on unknown short distance physics, be it gravity, GUTs or whatever.
We are therefore  fortunate that the issue can be finally subject to  experimental examination, made possible by the first  thorough exploration of the Fermi scale at the LHC. In any event, to know if the Fermi scale, or the Higgs boson mass, is natural or not will have very significant implications both of theoretical and of practical nature, influencing the strategy for future directions in particle physics at all. I think that the discovery of the Higgs-like particle at 125 GeV restricts the possibilities for a natural Fermi scale to two dominant cases: supersymmetry and the Higgs-as-a-pseudo-Goldstone-boson.

\begin{figure}[tb]
\centering
\includegraphics[width=.7\textwidth]{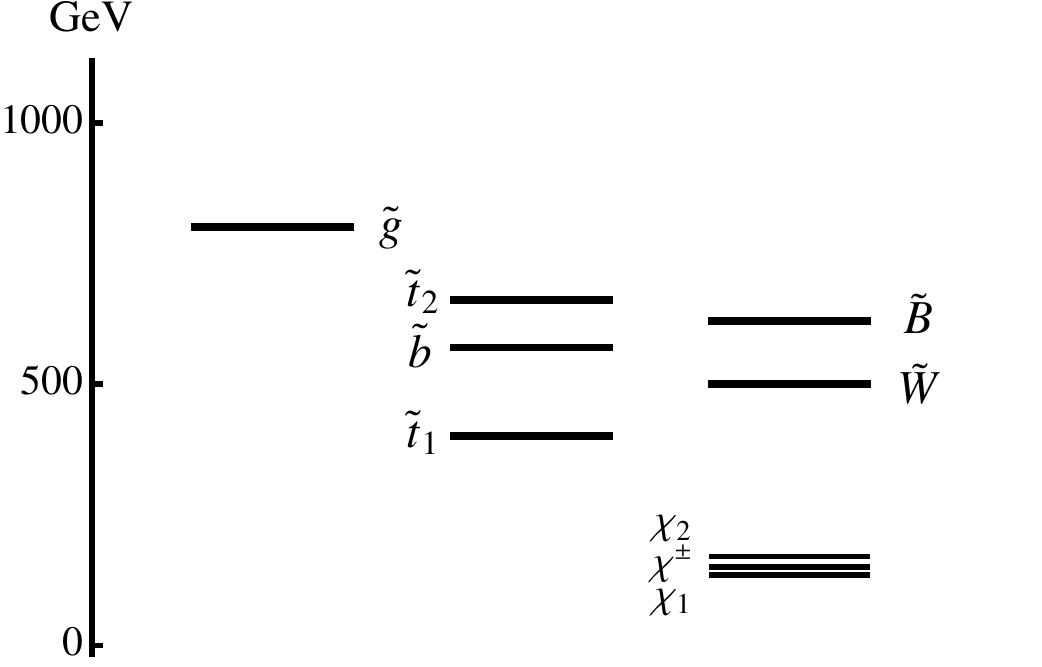}
\caption{A particular representation of a natural supersymmetric spectrum. Taken from  \cite{Barbieri:2009ev}
}
\label{nat_spectrum}
\end{figure}

\subsubsection{Supersymmetry \cite{Pomarol,Sundrum,Parker}}
\begin{figure}[tb]
\centering
\includegraphics[height=.7\textwidth]{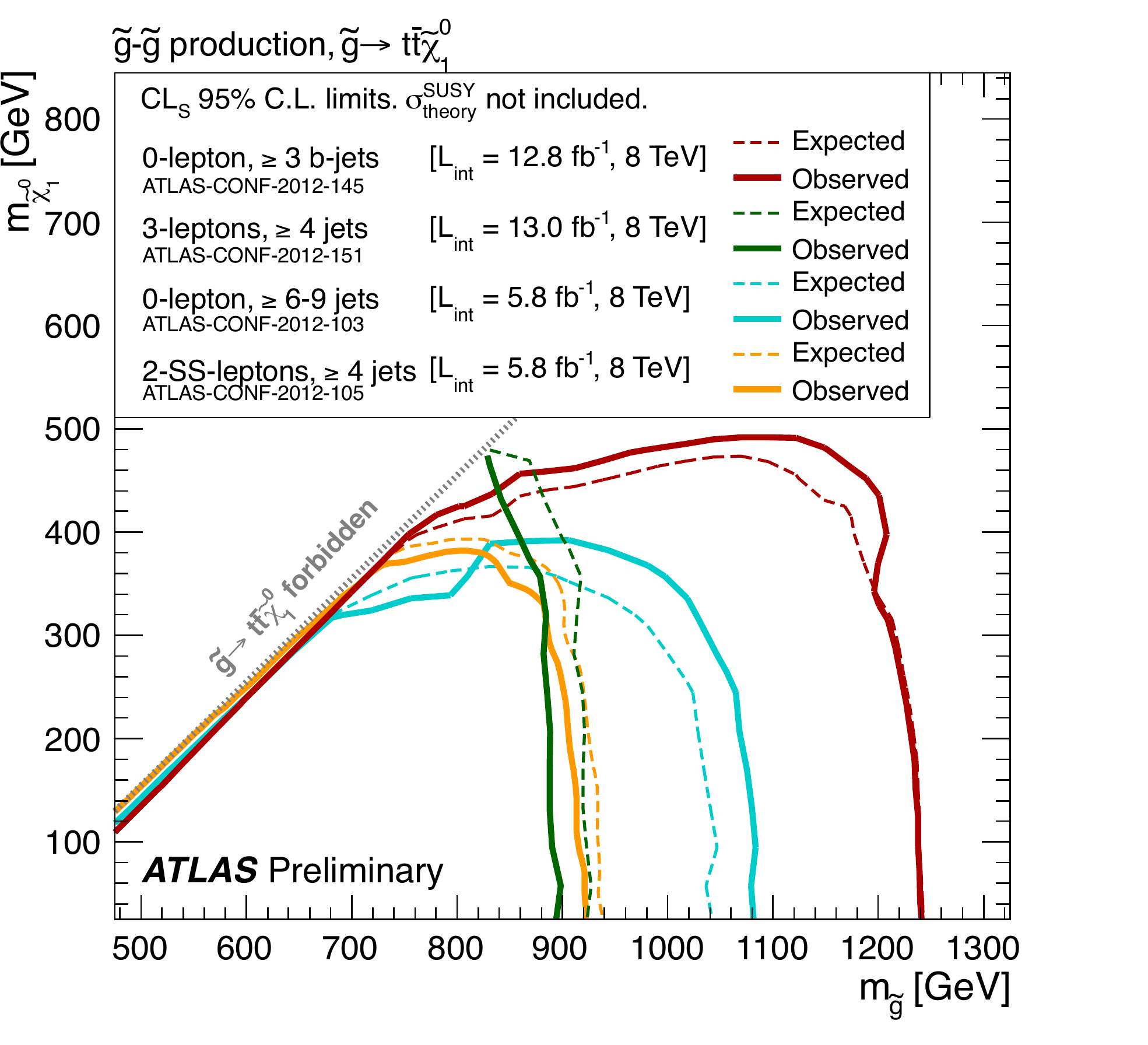}
\caption{Exclusion plot from $pp\rightarrow \tilde{g}~\tilde{g},~ \tilde{g}\rightarrow  t ~\bar{t}~\chi^0$. Taken from \cite{ATLAS}}
\label{nat_spectrum}
\end{figure}

%
%\begin{figure}[tb]
%\centering
%\includegraphics[width=\textwidth]{ATLAS_SUSY_Gtt_hcp12.pdf}
%\caption{Exclusion plot from $pp\rightarrow \tilde{g}~\tilde{g},~ \tilde{g}\rightarrow  t ~\bar{t}~\chi^0$. Taken from \cite{ATLAS}}
%\label{nat_spectrum}
%\end{figure}

LEP has shown no abundance of supersymmetric particles below the Z mass. Even before the start of the LHC, this has suggested focussing the attention of the searches to those particles that play a crucial role in the naturalness problem, like in the prototype spectrum of Fig. 1: the lightest particles, $\chi_{1,2}, \chi^\pm$, mostly higgsinos, whose mass directly influences at tree level the $Z$-mass; the two stops, $\tilde{t}_{1,2}$ and the left-handed sbottom, $\tilde{b}$, i.e. the s-particles with the strongest couplings to the Higgs boson; the gluino, $\tilde{g}$, which is coupled indirectly to the Higgs boson via its strong interactions  with the stops and the sbottom. The remaining particles shown in Fig. 1, mostly a wino, $\tilde{w}$, and a bino, $\tilde{b}$,  are not strongly constrained by naturalness but are generally expected to be lighter than the gluino. The heavier are these electroweak gauginos, the stronger is the degeneracy of the higgsinos.
Also not significantly constrained by naturalness are all the squarks and sleptons not represented in Fig. 1 and therefore allowed to be heavier. The $U(2)^3$ flavour symmetry mentioned in Section 4 could be related to this splitting between the third and  the first two generations of squarks, distinguished from it at some fundamental level.

\begin{figure}[tb]
\centering
\includegraphics[width=.9\textwidth]{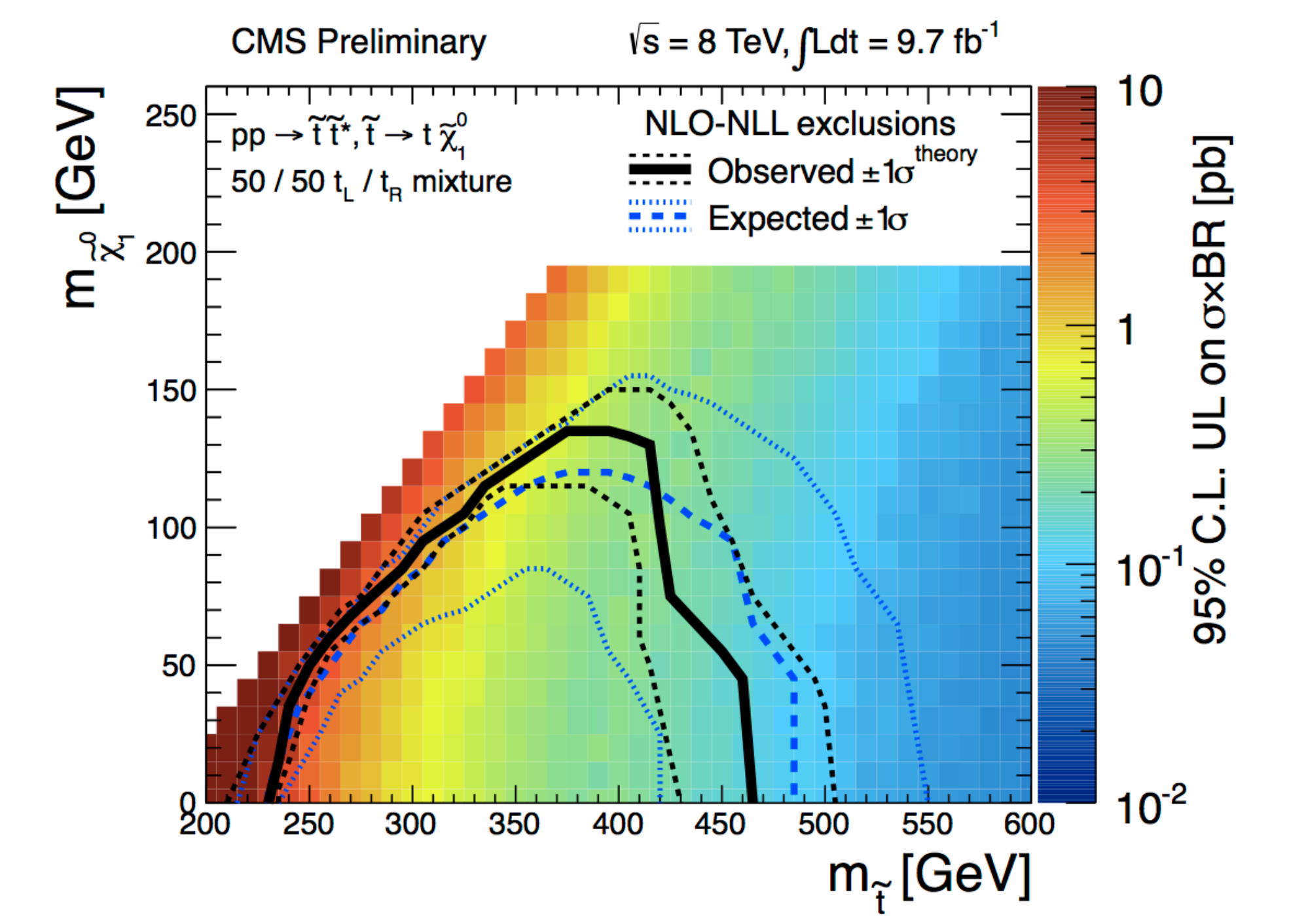}
\caption{Exclusion plot from $pp\rightarrow \tilde{t}~\tilde{t}^*,~ \tilde{t}\rightarrow  t \chi^0$. Taken from \cite{CMS}}
\label{nat_spectrum}
\end{figure}

The precise configuration of Fig. 1 is now excluded by the LHC searches, as visible from Fig. 2, preliminarly produced after ICHEP2012. Is this the end of natural supersymmetry? At least in part the answer to this question depends on where the quartic Higgs coupling comes from. If it is a pure $SU(2)_L\times U(1)$ gauge coupling, as in the Minimal Supersymmetric Standard Model, a  125 GeV Higgs boson is hardly compatible with a natural supersymmetric spectrum.
A way to take care of this problem is to admit the presence of a tree level Yukawa contribution to the quartic Higgs coupling, as in the NMSSM, in which case the naturalness constraints on the masses of the particles in Fig. 1 are relaxed. 
Speaking for myself, and knowing that  no absolute statement can be made on this issue, I would find natural supersymmetry in trouble if the search in Fig. 2 were extended to gluino masses of $1.5\div 1.8$ TeV or if the searches for direct stop production were pushed to explore all  the plane of Fig. 3, consistently with the kinematical constraints. 

Other ways  to keep natural supersymmetry alive or even fully alternative directions, like split supersymmetry, are discussed in \cite{Sundrum}. 

\subsubsection{A composite $\phi$ \cite{Pomarol}}

To establish the "elementary" nature, up to energies well above the Fermi scale, of the newly found particle would be a particularly striking feature of an already fundamental discovery. At least for this reason it is important to explore the alternative possibility that the new particle, while keeping its Higgs-like role, be a composite state from a new strong interaction. The Higgs-like particle, actually the entire $\phi$ multiplet,  might for example be itself a pseudo-Goldstone boson from a spontaneously broken global symmetry of the new putative strong dynamics. After a proper orientation of the vacuum, the scale $f$ of this breaking would have to be somewhat higher than the the Fermi scale itself, or the usual $v$ parameter. The standard $SU(2)_L\times U(1)$ gauge couplings and, even more so, the effective top Yukawa coupling to the composite Higgs boson are the sources of this last step of breaking.

 In analogy with the role of the stops in the case of supersymmetry, but also quite differently, what serves the purpose of keeping under control the biggest loop corrections to the Higgs squared mass in this case are new composite fermions, vector-like under the standard gauge interactions \cite{Redi}. Their typical mass is $m_\psi = Y f$, where $Y$ is a coupling constant that also normally controls the Higgs boson mass itself,
\begin{equation}\label{mh}
m_h = C \frac{\sqrt{N_c}}{\pi} m_t Y,
\end{equation}
where $N_c=3$ is the number of colors, $m_t$ is the top mass and $C$ is a model dependent coefficient of $\mathcal{O}(1)$. Barring unnatural fine-tunings, this very equation and the measured $m_h = 125$ GeV  call therefore for a semi-perturbative value of $Y$, i.e. for masses $m_\Psi$ below or at about 1 TeV, for minimal values of $f = 500 \div 800$ GeV. Not surprisingly, the direct search for these fermions is the key to test this picture. To respect the main constraints from the  EWPT, these coloured  top partners have to come in representations of a global $SU(2)_L\times SU(2)_R\times U(1)_X$ group, as such with several possible charges, $Q=5/3, 2/3, -1/3$. Current searches at LHC, limited to the data taken at $\sqrt{s} = 7$ TeV,  exclude $Q=5/3$ states below about 700 GeV. An interesting  question is to see how the composite Higgs picture copes with the  constraints from flavour and the full EWPT.

\subsection{What if the $\phi$ mass is not natural?}

\begin{figure}[t]
$$\includegraphics[height=5.35cm]{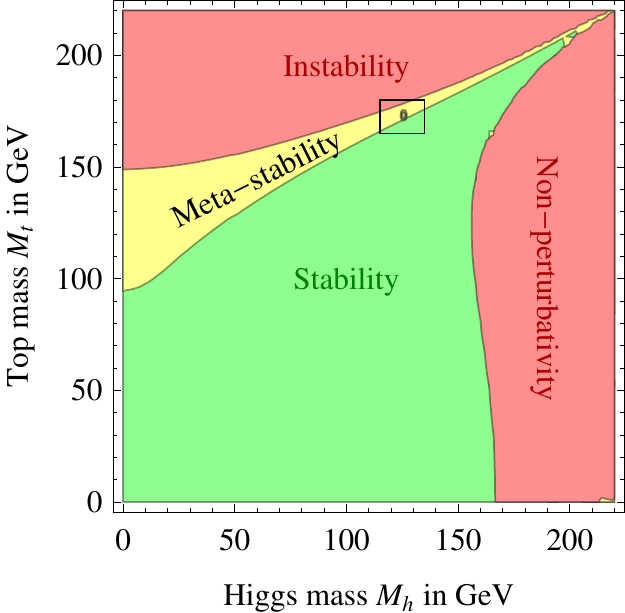}\qquad
\includegraphics[width=0.6\textwidth]{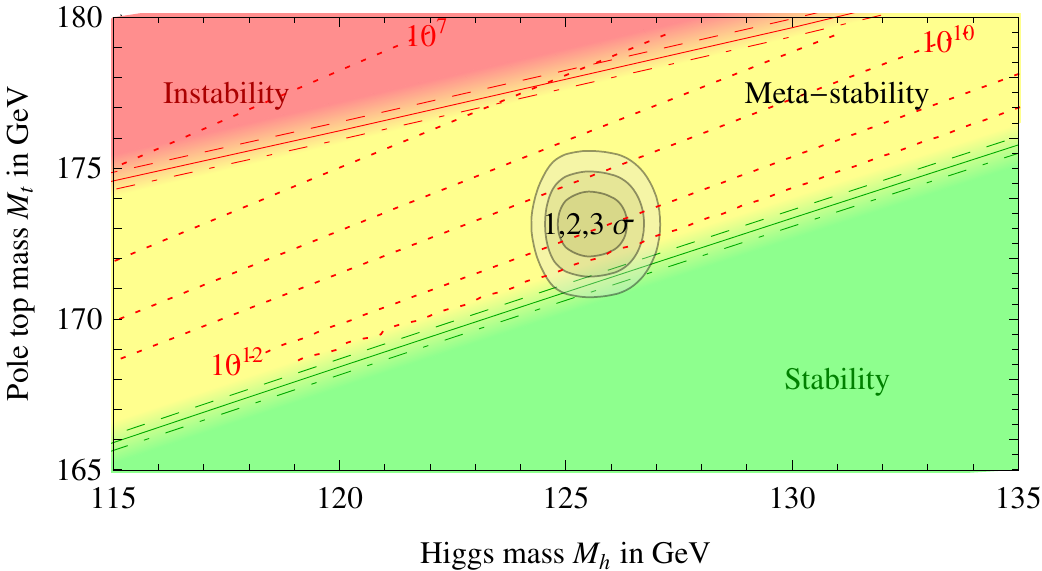}$$
\caption{Regions of absolute stability, meta-stability and instability of the SM vacuum 
in the $M_t$--$M_h$ plane (left), zoomed
in the region of the current experimental range of $M_h$ and $M_t$ (right). Taken from 
\cite{Degrassi:2012ry}}
\label{fig:regions}
\end{figure}

While the hypothesis of a natural Higgs boson is being put under severe experimental scrutiny, one has to contemplate the case that the Higgs boson is not natural at all. Given the   number of different possible directions that this opens up, the first question that can be asked is what happens if one assumes that the SM evolves unchanged up to very high energies. Recent accurate  calculations \cite{Chetyrkin:2012rz,Bezrukov:2012sa,Degrassi:2012ry,Alekhin:2012py} of the running of the various couplings  show that the Higgs boson self coupling, which relates at tree level to the Higgs mass as
\begin{equation}
\lambda = \frac{G_F m_h^2}{\sqrt{2}},
\end{equation}
starts at the Fermi scale at a value of about $0.1$ and slowly flattens out, crossing zero at a high energy scale quite dependent on the value of the top mass: $\mu \approx 10^{10}$ GeV for $m_{t} = 173.1$ GeV. To keep $\lambda$ above zero up to the Plank scale would require a low value of $m_t$,  about $3 \sigma$ away from the current central value $m_t = 173.1 \pm 0.7$ GeV. All this reflects itself in the location, shown in Fig. 4 left, of the measured values of $m_h$ and $m_t$ in the phase diagram of the SM Higgs potential \cite{Cabibbo:1979ay}, assumed not to be significantly distorted by gravity interactions at $|\phi|$ close to $M_{Pl}$.
The remarkable coincidence for which the SM appears to live right at the
border between the stability and instability regions motivates a zoom of the relevant spot in Fig. 4  right.  The dotted lines show the instability scale in GeV. Again stability at $M_{Pl}$ is shown to be reached at   low  values of $m_t$ by $2\div 3 \sigma$.
The possible existence of a second degenerate minimum of the SM Higgs potential at $\phi$ near $M_{Pl}$ has been suggested long ago \cite{Bennett:1988xi} and discussed again more recently in different contexts, among which are \cite{Bezrukov:2009db,Hall:2009nd,
Shaposhnikov:2009pv,Masina:2011aa}. Suppose that refined measurements  of the top mass, with suitable theory support, allowed this possibility. A non rhetoric question is: how to know that this is not a coincidence? If the Higgs mass is unnatural, it is conceivable that we shall be faced with more than one question of this kind.

\section{Aknowledgements}

It is pleasure to thank the organizers  of ICHEP2012. I am indebted to Enrico Bertuzzo, Dario Buttazzo, Marco Farina, Lawrence Hall, Gino Isidori, Paolo Lodone, Yasunori Nomura, Duccio Pappadopulo, Slava Rychkov, Filippo Sala, David Straub and Andrea Tesi for many useful discussions concerning the topics of this talk.
This work was supported by the EU ITN ``Unification in the LHC Era'', 
contract PITN-GA-2009-237920 (UNILHC) and by MIUR under contract 2010YJ2NYW-010


\begin{thebibliography}{99}

%\cite{:2012gk}
\bibitem{:2012gk}
  G.~Aad {\it et al.}  [ATLAS Collaboration],
  %``Observation of a new particle in the search for the Standard Model Higgs boson with the ATLAS detector at the LHC,''
  Phys.\ Lett.\ B {\bf 716} (2012) 1
  [arXiv:1207.7214 [hep-ex]].
  %%CITATION = ARXIV:1207.7214;%%
  %\cite{:2012gu}
\bibitem{:2012gu}
  S.~Chatrchyan {\it et al.}  [CMS Collaboration],
  %``Observation of a new boson at a mass of 125 GeV with the CMS experiment at the LHC,''
  Phys.\ Lett.\ B {\bf 716} (2012) 30
  [arXiv:1207.7235 [hep-ex]].
  %%CITATION = ARXIV:1207.7235;%%
\bibitem{Dixon} 
L. Dixon, these Proceedings, and references therein.
\bibitem{Campbell}
 J. Campbell, these Proceedings, and references therein.
\bibitem{Kosower} 
 D. Kosower, these Proceedings, and references therein.
\bibitem{Kobayashi} 
T. Kobayashi, these Proceedings, and references therein.
\bibitem{Cao} 
J. Cao, these Proceedings, and references therein.
\bibitem{Gonzalez} 
C. Gonzalez-Garcia, these Proceedings, and references therein.
\bibitem{Kuno} 
J. Kuno, these Proceedings, and references therein.
\bibitem{Perez} 
G. Perez, these Proceedings, and references therein.
\bibitem{Stone} 
S. Stone, these Proceedings, and references therein.
\bibitem{Tarantino} 
C. Tarantino, these Proceedings, and references therein.
\bibitem{Tonelli} 
D. Tonelli, these Proceedings, and references therein.
\bibitem{Garra} 
J. Garra Tico, these Proceedings, and references therein.

\bibitem{Soni} 
A. Soni, these Proceedings, and references therein.
\bibitem{DeNardo} 
G. De Nardo, these Proceedings, and references therein.
\bibitem{Yook} 
Y. Yook, these Proceedings, and references therein.
\bibitem{Nakao} 
M. Nakao, these Proceedings, and references therein.
\bibitem{Urquijo} 
P. Urquijo, these Proceedings, and references therein.
\bibitem{Grohsjean} 
A. Grohsjean, these Proceedings, and references therein.
\bibitem{Hays} 
C. Hays, these Proceedings, and references therein.
%\cite{Englert:1964et}
\bibitem{Englert:1964et}
  F.~Englert and R.~Brout,
  %``Broken Symmetry and the Mass of Gauge Vector Mesons,''
  Phys.\ Rev.\ Lett.\  {\bf 13} (1964) 321.
  %%CITATION = PRLTA,13,321;%%
  %\cite{Higgs:1964pj}
\bibitem{Higgs:1964pj}
  P.~W.~Higgs,
  %``Broken Symmetries and the Masses of Gauge Bosons,''
  Phys.\ Rev.\ Lett.\  {\bf 13} (1964) 508.
  %%CITATION = PRLTA,13,508;%%

%\cite{Baak:2012kk}
\bibitem{Baak:2012kk}
  M.~Baak, M.~Goebel, J.~Haller, A.~Hoecker, D.~Kennedy, R.~Kogler, K.~Moenig and M.~Schott {\it et al.},
  %``The Electroweak Fit of the Standard Model after the Discovery of a New Boson at the LHC,''
  Eur.\ Phys.\ J.\ C {\bf 72} (2012) 2205
  [arXiv:1209.2716 [hep-ph]].
  %%CITATION = ARXIV:1209.2716;%%

\bibitem{Hawkings} 
R. Hawkings, these Proceedings, and references therein.
\bibitem{Incandela} 
J. Incandela, these Proceedings, and references therein.
\bibitem{Shalhout} 
S. Shalhout et al, these Proceedings, and references therein.
\bibitem{Pomarol} 
A. Pomarol, these Proceedings, and references therein.
\bibitem{Sundrum} 
R. Sundrum, these Proceedings, and references therein.
\bibitem{Parker} 
A. Parker, these Proceedings, and references therein.

\bibitem{Barbieri:2009ev}
  R.~Barbieri and D.~Pappadopulo,
  %``S-particles at their naturalness limits,''
  JHEP {\bf 0910} (2009) 061
  [arXiv:0906.4546 [hep-ph]].


\bibitem{ATLAS}
 ATLAS Collaboration, presented at HCP2012
 \bibitem{CMS}
 CMS Collaboration, SUS-12-023
 %\cite{Chetyrkin:2012rz}
 \bibitem{Redi} 
M. Redi, these Proceedings, and references therein.
\bibitem{Chetyrkin:2012rz}
  K.~G.~Chetyrkin and M.~F.~Zoller,
  %``Three-loop \beta-functions for top-Yukawa and the Higgs self-interaction in the Standard Model,''
  JHEP {\bf 1206} (2012) 033
  [arXiv:1205.2892 [hep-ph]].
  %%CITATION = ARXIV:1205.2892;%%
  %\cite{Bezrukov:2012sa}
\bibitem{Bezrukov:2012sa}
  F.~Bezrukov, M.~Y.~.Kalmykov, B.~A.~Kniehl and M.~Shaposhnikov,
  %``Higgs Boson Mass and New Physics,''
  JHEP {\bf 1210} (2012) 140
  [arXiv:1205.2893 [hep-ph]].
  %%CITATION = ARXIV:1205.2893;%%
  %\cite{Degrassi:2012ry}
\bibitem{Degrassi:2012ry}
  G.~Degrassi, S.~Di Vita, J.~Elias-Miro, J.~R.~Espinosa, G.~F.~Giudice, G.~Isidori and A.~Strumia,
  %``Higgs mass and vacuum stability in the Standard Model at NNLO,''
  JHEP {\bf 1208} (2012) 098
  [arXiv:1205.6497 [hep-ph]].
  %%CITATION = ARXIV:1205.6497;%%
  %\cite{Alekhin:2012py}
\bibitem{Alekhin:2012py}
  S.~Alekhin, A.~Djouadi and S.~Moch,
  %``The top quark and Higgs boson masses and the stability of the electroweak vacuum,''
  Phys.\ Lett.\ B {\bf 716} (2012) 214
  [arXiv:1207.0980 [hep-ph]].
  %%CITATION = ARXIV:1207.0980;%%
  %\cite{Cabibbo:1979ay}
\bibitem{Cabibbo:1979ay}
  N.~Cabibbo, L.~Maiani, G.~Parisi and R.~Petronzio,
  %``Bounds on the Fermions and Higgs Boson Masses in Grand Unified Theories,''
  Nucl.\ Phys.\ B {\bf 158} (1979) 295.
  %%CITATION = NUPHA,B158,295;%%
 
 %\cite{Bennett:1988xi}
\bibitem{Bennett:1988xi}
  D.~L.~Bennett, H.~B.~Nielsen and I.~Picek,
  %``Understanding Fine Structure Constants And Three Generations,''
  Phys.\ Lett.\ B {\bf 208} (1988) 275.
  %%CITATION = PHLTA,B208,275;%%
  
  %\cite{Bezrukov:2009db}
\bibitem{Bezrukov:2009db}
  F.~Bezrukov and M.~Shaposhnikov,
  %``Standard Model Higgs boson mass from inflation: Two loop analysis,''
  JHEP {\bf 0907} (2009) 089
  [arXiv:0904.1537 [hep-ph]].
  %%CITATION = ARXIV:0904.1537;%%
  %\cite{Hall:2009nd}
\bibitem{Hall:2009nd}
  L.~J.~Hall and Y.~Nomura,
  %``A Finely-Predicted Higgs Boson Mass from A Finely-Tuned Weak Scale,''
  JHEP {\bf 1003} (2010) 076
  [arXiv:0910.2235 [hep-ph]].
  %%CITATION = ARXIV:0910.2235;%%
  %\cite{Shaposhnikov:2009pv}
\bibitem{Shaposhnikov:2009pv}
  M.~Shaposhnikov and C.~Wetterich,
  %``Asymptotic safety of gravity and the Higgs boson mass,''
  Phys.\ Lett.\ B {\bf 683} (2010) 196
  [arXiv:0912.0208 [hep-th]].
  %%CITATION = ARXIV:0912.0208;%%
  %\cite{Masina:2011aa}
\bibitem{Masina:2011aa}
  I.~Masina and A.~Notari,
  %``The Higgs mass range from Standard Model false vacuum Inflation in scalar-tensor gravity,''
  Phys.\ Rev.\ D {\bf 85} (2012) 123506
  [arXiv:1112.2659 [hep-ph]].
  %%CITATION = ARXIV:1112.2659;%%

\end{thebibliography}
\end{document}